\documentstyle[12pt,psfig]{article}

\catcode`\@=11
\long\def\@makefntext#1{
\protect\noindent \hbox to 3.2pt {\hskip-.9pt
$^{{\ninerm\@thefnmark}}$\hfil}#1\hfill}                

\def\@makefnmark{\hbox to 0pt{$^{\@thefnmark}$\hss}}  

\def\ps@myheadings{\let\@mkboth\@gobbletwo
\def\@oddhead{\hbox{}
\rightmark\hfil\ninerm\thepage}
\def\@oddfoot{}\def\@evenhead{\ninerm\thepage\hfil
\leftmark\hbox{}}\def\@evenfoot{}
\def\sectionmark##1{}\def\subsectionmark##1{}}

\renewcommand{\thefootnote}{\fnsymbol{footnote}}

\def\sectionc{\@startsection {section}{1}{\z@}{-3.5ex plus -1ex minus
    -.2ex}{2.3ex plus .2ex}{\bf }}
\def\subsectionc{\@startsection{subsection}{2}{\z@}{-3.25ex plus -1ex minus
   -.2ex}{1.5ex plus .2ex}{\it }}
\renewcommand{\section}[1]{\sectionc{#1}\hspace*{\parindent}}
\renewcommand{\subsection}[1]{\subsectionc{#1}\hspace*{\parindent}}
\newcounter{appendixc}
\newcounter{subappendixc}[appendixc]
\newcounter{subsubappendixc}[subappendixc]

\renewcommand{\appendix}[1] {\vspace*{0.6cm}
        \refstepcounter{appendixc}
        \setcounter{figure}{0}
        \setcounter{table}{0}
        \setcounter{equation}{0}
        \renewcommand{\thefigure}{\Alph{appendixc}.\arabic{figure}}
        \renewcommand{\thetable}{\Alph{appendixc}.\arabic{table}}
        \renewcommand{\theappendixc}{\Alph{appendixc}}
        \renewcommand{\theequation}{\Alph{appendixc}.\arabic{equation}}
        \noindent{\bf Appendix \theappendixc #1}\par\vspace*{0.4cm}}

\def\abstracts#1{{

\centering{\begin{minipage}{13.2truecm}
\footnotesize\baselineskip=13pt\noindent
        \parindent=0pt #1
        \end{minipage}}\par}}

\renewenvironment{thebibliography}[1]
        {\begin{list}{\arabic{enumi}.}
        {\usecounter{enumi}\setlength{\parsep}{0pt}
\setlength{\leftmargin 0.75cm}{\rightmargin 0pt}
         \setlength{\itemsep}{0pt} \settowidth
        {\labelwidth}{#1.}\sloppy}}{\end{list}}

\newcounter{itemlistc}
\newcounter{romanlistc}
\newcounter{alphlistc}
\newcounter{arabiclistc}

\newcommand{\fcaption}[1]{
        \refstepcounter{figure}
        \setbox\@tempboxa = \hbox{\footnotesize Figure~\thefigure. #1}
        \ifdim \wd\@tempboxa > 6in
           {\begin{center}
        \parbox{6in}{\footnotesize\baselineskip=13pt Figure~\thefigure. #1}
            \end{center}}
        \else
             {\begin{center}
             {\footnotesize Figure~\thefigure. #1}
              \end{center}}
        \fi}

\newcommand{\tcaption}[1]{
        \refstepcounter{table}
        \setbox\@tempboxa = \hbox{\footnotesize Table~\thetable. #1}
        \ifdim \wd\@tempboxa > 6in
           {\begin{center}
        \parbox{6in}{\footnotesize\baselineskip=13pt Table~\thetable. #1}
            \end{center}}
        \else
             {\begin{center}
             {\footnotesize Table~\thetable. #1}
              \end{center}}
        \fi}

\def\@citex[#1]#2{\if@filesw\immediate\write\@auxout
        {\string\citation{#2}}\fi
\def\@citea{}\@cite{\@for\@citeb:=#2\do
        {\@citea\def\@citea{,}\@ifundefined
        {b@\@citeb}{{\bf ?}\@warning
        {Citation `\@citeb' on page \thepage \space undefined}}
        {\csname b@\@citeb\endcsname}}}{#1}}

\newif\if@cghi
\def\cite{\@cghitrue\@ifnextchar [{\@tempswatrue
        \@citex}{\@tempswafalse\@citex[]}}
\def\citelow{\@cghifalse\@ifnextchar [{\@tempswatrue
        \@citex}{\@tempswafalse\@citex[]}}
\def\@cite#1#2{{$\null^{#1}$\if@tempswa\typeout
        {IJCGA warning: optional citation argument
        ignored: `#2'} \fi}}

 1
 1
 1

\font\ninerm=cmr9

\begin{document}

\centerline{\normalsize\bf THEORY OF ETA PHOTO- AND 
ELECTROPRODUCTION\footnote{Presented by Nimai C. Mukhopadhyay.
Invited talk at the CEBAF/INT Workshop on N$^*$ Physics, Sept. 9-13, 1996.}}
\baselineskip=15pt

\vspace*{0.6cm}
\centerline{\footnotesize NIMAI C. MUKHOPADHYAY, J. -F. ZHANG}
\baselineskip=13pt
\centerline{\footnotesize\it Department of Physics, Applied Physics 
and Astronomy}
\baselineskip=13pt
\centerline{\footnotesize\it Rensselaer Polytechnic Institute, Troy, NY 
12180-3590, U.S.A.}
\vspace*{0.3cm}
\vspace*{0.3cm}
\centerline{\footnotesize M. BENMERROUCHE}
\baselineskip=13pt
\centerline{\footnotesize\it Saskatchewan Accelerator Laboratory,
University of Saskatchewan} 
\centerline{\footnotesize\it Saskatoon, Canada SK S7N 5C6}

\vspace*{0.6cm}
\abstracts{We analyze the available data on eta photo- and
electroproduction, around $W\approx 1535 MeV$, in the framework of
the effective Lagrangian approach, and extract, in a nearly 
model-independent fashion, the electrostrong amplitude for the
$\gamma N\rightarrow N^{*}(1535)\rightarrow N\eta$ processes. Quark model
approaches are shown to be {\it quite inadequate} to explain this
property at {\it all} $Q^2$. In  particular, at {\it high} $Q^2$,
the extracted amplitude falls {\it much slower} than the predictions
of the quark model, as a function of $Q^2$, a situation similar
to the electroexcitation and decay of $\Delta$(1232). A QCD 
explanation of these observations is urgently needed.}

\normalsize\baselineskip=15pt
\setcounter{footnote}{0}
\renewcommand{\thefootnote}{\alph{footnote}}

\section{Introduction}\label{sec:intro}
Nathan Isgur, in his introductory talk\cite{ref1}, has emphasized
the importance of studying baryons in the overall context of 
studying QCD: baryons are among the basic asymptotic states of
QCD, having manifest non-Abelian physics and being essential in our
attempt to understand nuclear physics in the QCD context. While QCD is 
a theory of quarks and gluons, quark models are attempts 
to use the effective degrees of freedom of valence quarks, with
remarkable relationships to the quenched approximation on the
lattice\cite{ref2}. Thus, rigorous tests of quark models are
useful towards our understanding of QCD. Our work reported here
should  aid in this effort by providing one piece of important
physics: that of the electromagnetic excitation of the N$^*$(1535)
resonance and its strong decay to the N$\eta$ channel. The
processes\cite{ref3,ref4,ref5,ref6,ref7,ref8,ref9}
\begin{equation}
\gamma +p\rightarrow p+\eta ,
\end{equation}
\begin{equation}
\gamma +n\rightarrow n+\eta,
\end{equation}
where the photon $\gamma$ is real or virtual, are powerful tools
to study the N$^*$(1535) resonance, due to the special property of
N$^*$(1535)  coupling strongly  to the N$\eta$ channel (N, nucleon), as
opposed to many other N$^*$'s which do not have significant
couplings to this channel\cite{ref10}. We shall use an effective 
Lagrangian approach (ELA), with the usual Born terms, vector
meson exchanges and N$^*$ excitations\cite{ref3}[Fig.1], and
treat the existing cross section\cite{ref5,ref8,ref9} and
 polarization data\cite{ref6}. Our output of these analyses
will be a physical (dressed) parameter for the excitation of
the N$^*$(1535) resonance and its decay as a function of $Q^2$,
negative of the momentum transfer squared. In comparing with
other calculations, only dressed or physical parameters need be
compared. As discussed in the classical text on quantum
electrodynamics\cite{ref11}, the relation between ``bare" and 
physical (or ``dressed") parameters is always model-dependent\cite{ref12}.

\begin{figure}[t]
\begin{center}
\vskip 1.0cm
\psfig{figure=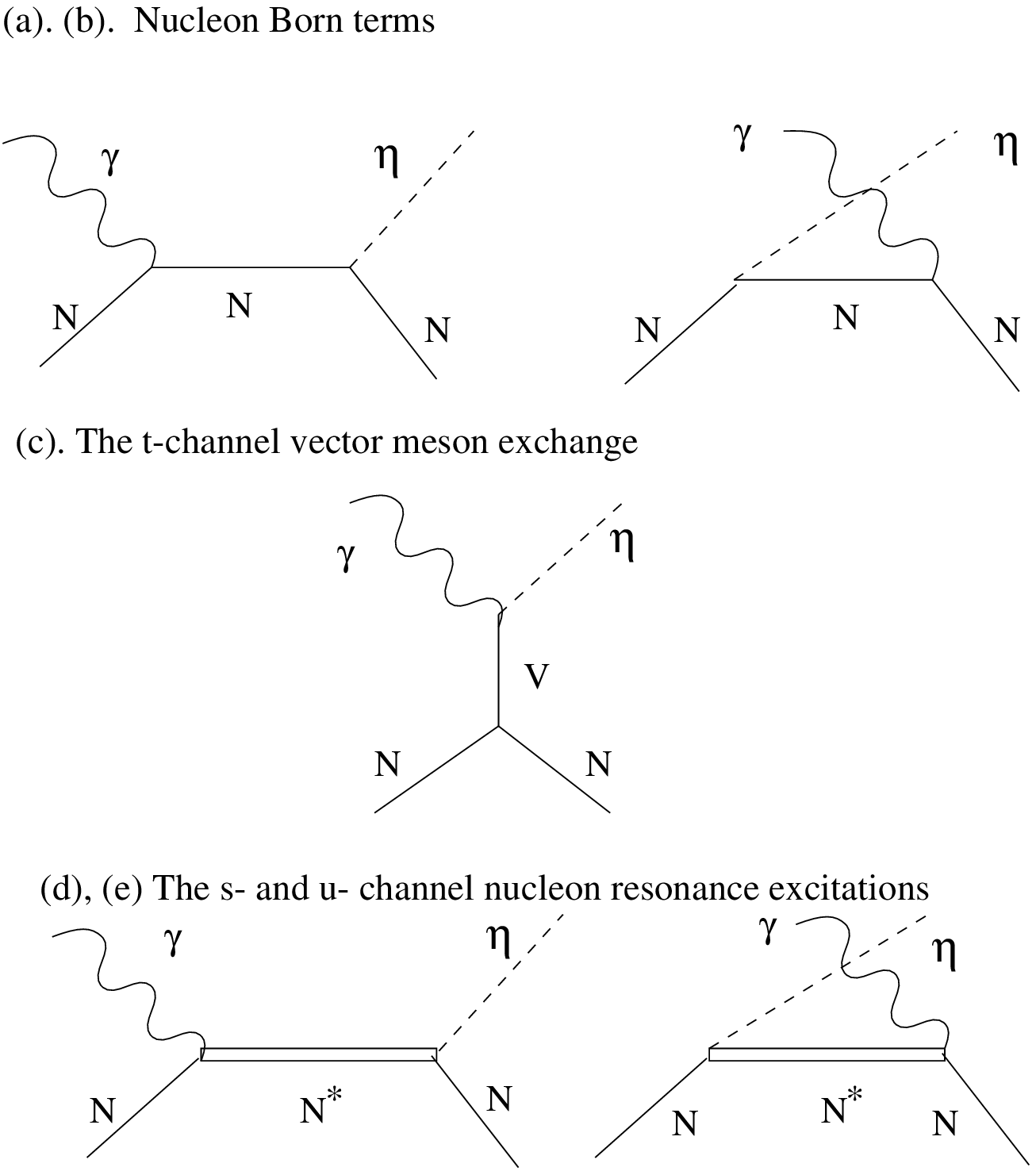,height=4.5in}
\end{center}
\fcaption{Mechanisms of Photo- and electroproduction of eta mesons
in our approach\protect{\cite{ref3}}
\label{fig:fy}}
\end{figure}

We recall at the outset that the quark shell model and lattice gauge
calculations agree surprisingly well in the case of the
 $\gamma N\rightarrow \Delta$
transition. The magnetic dipole amplitude (M1) that is 
extracted\cite{ref13} from the data (old or new)\cite{ref14}
is about 285 to 290 units, while the quark model\cite{ref15} or
the lattice calculation\cite{ref2} is about 210 units (see Table I).
Thus, there seems to be a significant transition magnetism
shortage in these  calculations. Could it be due to the 
$q\bar{q}$ pair effects, discussed by Isgur\cite{ref1}? We will see
in the future investigations. Suffice to say that this magnetism
shortage is confirmed by the Compton scattering as well\cite{ref16}.
At high $Q^2$, the data seem to scale and even show hints of 
$logs$\cite{ref17}, a situation {\it very different} from the
quark model anticipations. Even in the model of Salme 
{\it et al.}\cite{ref15},
it would be difficult to reproduce all these features.

\begin{table}[t]
\protect
\tcaption{A summary of the $N\rightarrow \Delta$ magnetic dipole
amplitude (in conventional units)}\label{tab:sum}
\small
\vspace{0.4cm}
\begin{center}
\begin{tabular}{|c|l|c|}
\hline
 & & \\
Method&
Authors &Result
\\ \hline
K-matrix residue&Davidson,Mukhopadhyay\protect\cite{ref13} &$290\pm
13$\\ \hline
ELA fits&Davidson, Mukhopadhyay and Wittman\protect\cite{ref13} &
$285\pm 37$ \\ \hline
Quark model &Koniuk et al.\protect\cite{ref15}& $206$
\\ \hline
Lattice & Leinweber, Draper and Woloshyn\protect\cite{ref2}& $ 210\pm 25$
\\ 
 & &  \\ \hline
\end{tabular}
\end{center}
\end{table}

Remainder of this paper will be the following: in the next section,
we survey the current literature to provide a quick review of
the current excitements on N$^*$(1535). 
Section 3 contains an  overview of the
ELA. Section 4 gives our basic results of the ELA for 
photoproduction. Section 5 does the same for electroproduction.
Section 6 summarizes our conclusions.

\section{Some recent excitements on N$^*$(1535)}\label{sec:some}
On the experimental side, we have seen at this workshop the
beautiful high quality data on eta photoproduction from Mainz in
Krusche's talk\cite{ref5}. Other new data\cite{ref5}
 from Bates and Bonn
do not compare in quality with these ones. Some new polarization 
data\cite{ref7} are also coming from Bonn and GRAAL, but we still
have to use the old data base\cite{ref6}. For electroproduction
of the eta mesons, the cross section data are all old\cite{ref9},
except for one set coming from the Elan collaboration\cite{ref18}
at the ELSA
ring, at $Q^2=0.056 GeV^2$. However, these recent data show
rather strange angular distributions.

On the theory side, the first recent examinations of the ELA were
done by the RPI group\cite{ref3}. Bennhold and Tanabe\cite{ref19}
and Sauermann {\it et al.}\cite{ref19} have emphasized 
coupled-channel approaches. General phenomenology has been reviewed
by Kn\"{o}chlein {\it et al.}\cite{ref19}.

For polarization observables, the nodal trajectory and its
value in illuminating the amplitude structure has been stressed by
Saghai, Tabakin and collaborators\cite{ref20}.

On the quark model, the works of Koniuk and Isgur\cite{ref15},
Close and Li\cite{ref15}, Weber\cite{ref15} and Capstick\cite{ref15}
have set the stage for more  investigations. In this conference,
we have had different approaches of quark model presented by
Keister\cite{ref15}, Salme\cite{ref15}, Iachello\cite{ref15} and
Leviatan\cite{ref15}. In particular, questions have been raised as to
whether N$^*$(1535) is a conventional $q^3$ state\cite{ref21}. If 
the answer is no, where is the regular $q^3$ state with the same
quantum numbers? Thus, the processes (1) and (2) should address
some of these vital questions.

Whole sets of issues come up in discussing the relationships
between $\eta$ and $\eta^{\prime}$ that we have discussed
elsewhere\cite{ref22}. These include the chiral U(1) problem\cite{ref23},
the quark contents\cite{ref24} of $\eta$, $\eta^{\prime}$, the
$\eta_{1}$-$\eta_{8}$ mixing angle\cite{ref25} and so on.
Finally, Leinweber {\it et al.}\cite{ref2} have approached the
$\gamma NN^*$ amplitudes on the lattice,
while Oka and collaborators\cite{ref26} have compared the 
$\pi NN^*$ vs. $\eta NN^*$ coupling constants in the QCD sum
rules, indicating how the former can be suppressed relative to
the latter.

All in all, the subject of N$^*$(1535) is hot both theoretically
and experimentally. At TJNAF, there are experimental proposals
pending in {\it all} Halls of CEBAF.

\section{The Effective Lagrangian Approach (ELA)}\label{sec:ela}
We have discussed in detail the tree-level ELA in the 
literature\cite{ref3}. We shall quote the main conclusions of
these discussions here. First, the pseudoscalar eta-nucleon
coupling constant is {\it not} well-known. One can give a rather broad
range for $g_{\eta}$:
\begin{equation}
0.2\leq g_{\eta} \leq 6.2.
\end{equation}
The vector meson sector have effective strong parameters $g^{v}_{i}$
given by
\begin{eqnarray}
\lambda_{\rho}g^{\rho}_{v}+\lambda_{\omega}g_{v}^{\omega}&=&
5.93\pm 0.82,\\ \nonumber
\lambda_{\rho}g^{\rho}_{t}+\lambda_{\omega}g_{t}^{\omega}&=&
17.50\pm 2.57,
\end{eqnarray}
while the electromagnetic decay amplitudes $\lambda_{i}$ for
the vector mesons controlling the vector meson radiative
decay width are
\begin{eqnarray}
\lambda_{\rho}&=& 1.06\pm 0.15,\\ \nonumber
\lambda_{\omega}&=& 0.31\pm 0.06.
\end{eqnarray}
From our earlier experiences, we know the non-resonant Born sectors 
to be relatively unimportant. The main player is the s-channel
excitation of the N$^*$(1535), followed by some modest
importance of the excitation of the N$^*$(1520). The effective
Lagrangian involving the N$^*$(1535) is discussed here for brevity.
It is given by
\begin{equation}
L^{ps}_{\eta NR}=-ig_{\eta NR}\bar{N}R\eta +h.c.,
\end{equation}
\begin{equation}
L_{\gamma NR}=\frac{e}{2(M_{R}+M)}\bar{R}(k_{R}^{s}+k^{v}_{R}\tau_{3})
\gamma_{5}\sigma_{\mu\nu}NF^{\mu\nu}+h.c.,
\end{equation}
thereby introducing an effective parameter $\chi_{p}=k_{R}^{p}
g_{\eta pR}$, where $k_{R}^{p}=k_{R}^{s}+k_{R}^{v}$ for
proton targets and $\chi_{n}$, an appropriate parameter for 
neutron, describing the resonance dominated eta photoproduction.
For the electroproduction, we study the $q^2$ dependence of these
parameters, which allows us a rigorous test of the QCD-inspired
models, and ultimately QCD itself.

For the spin-3/2 (and higher spin) resonances, the effective
Lagrangian is more complicated, as we have discussed elsewhere\cite{ref27}.
The strong interaction Lagrangian, $L_{\eta NR}$, to
excite the resonance $R$, is given by \cite{ref27}
\begin{equation}
L_{\eta NR}=\frac{f_{\eta NR}}{\mu}\bar{R}^{\mu}\theta_{\mu\nu}(Z)
\gamma_{5}N\partial^{\nu}\eta +h.c.,
\end{equation}
where the term $\theta_{\mu\nu}(V)$ is given by
\begin{equation}
\theta_{\mu\nu}(V)=g_{\mu\nu}+[\frac{1}{2}(1+4V)A+V]\gamma_{\mu}\gamma_{
\nu}.
\end{equation}
The parameter $V(=Z$ here) is unknown, and has to be fitted as
an ``off-shell" parameter (two more would
come from the photon vertices for real photons), while 
the ``point-transformation"
 parameter $A$ drops out from observables. The above Lagrangian yields
an ``off-shell" spin-1/2 sector\cite{ref27}, influencing the non-resonant
multipoles. 

\section{Our Results for Real Photons}
We first summarize our results for the  photoproduction. In 
Fig.2, we show our beautiful fits of the Mainz angular distribution
data off proton at $E_{\gamma}=716$ and 790 MeV. In the former, the 
angular distribution is flat, indicating the dominance of the
$E_{0^+}$ multipoles, while the latter demonstrates the effects of
the higher partial waves.

\begin{figure}[t]
\begin{center}
\vskip 1.0cm
\psfig{figure=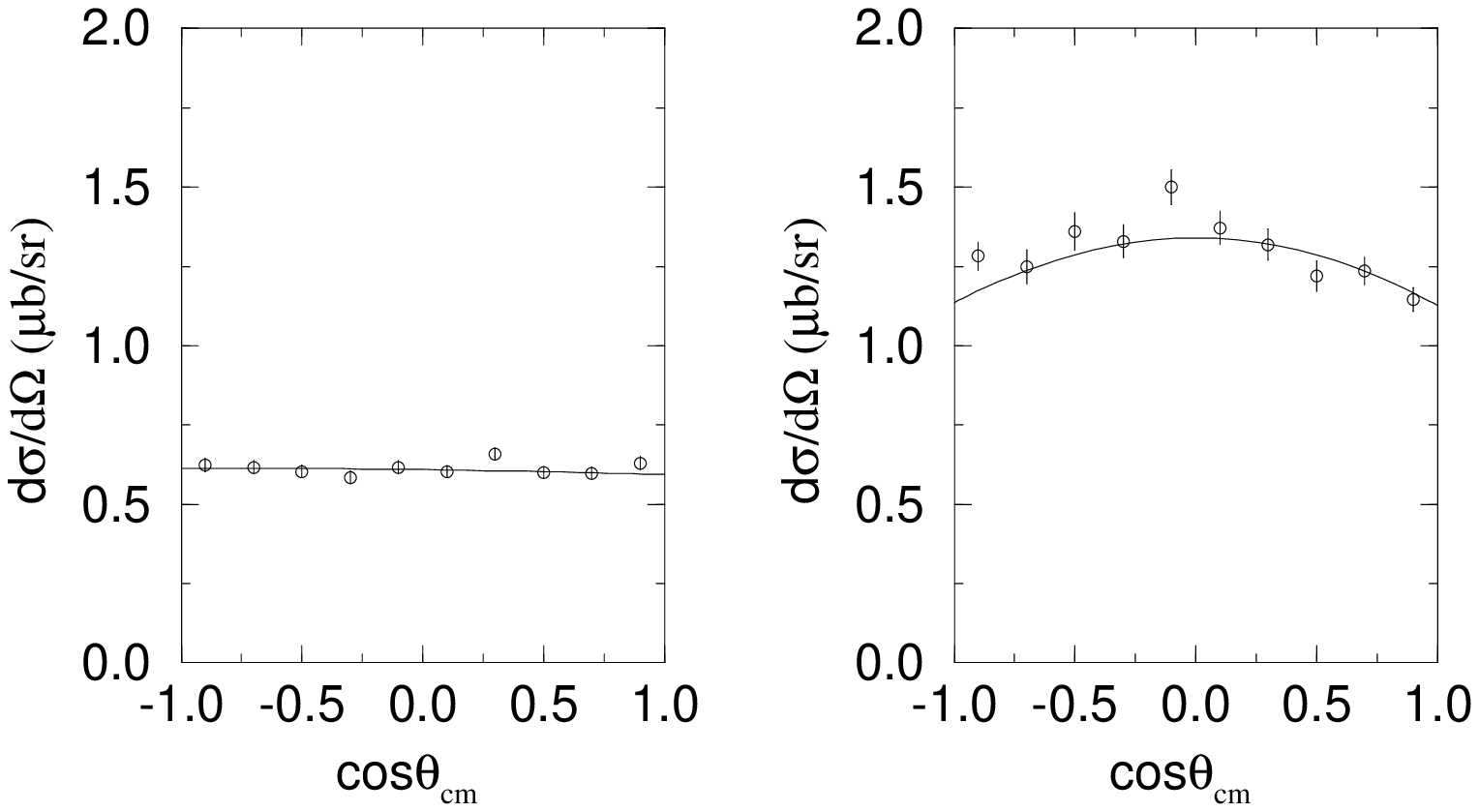,height=2.5in}
\end{center}
\fcaption{Angular distributions\protect\cite{ref3}
 at $E_{\gamma}=716$ and 790
MeV fitted by our ELA. Data from the Mainz experiment.\protect\cite{ref5}
\label{fig:fit}}
\end{figure}

We can extract from the above fits a nearly model independent
electrostrong parameter, related to $\chi_{p}$ and $\chi_{n}$
mentioned earlier\cite{ref3}:
\begin{equation}
\xi_{i}=\sqrt{(\zeta_{i}\Gamma_{\eta})}A_{1/2}^{i}/\Gamma_{T}
\end{equation}
where $i=n$, $p$, $\Gamma_{\eta}$ and $\Gamma_{T}$ are the partial
$\eta$N and total widths of the resonance N$^*$(1535), $\zeta_{i}$,
some kinemetic factor ($\zeta_{p}\approx\zeta_{n}\approx 1.6$),
 $A_{1/2}^{i}$, the photon helicity
amplitude. It is this parameter that needs to be computed in QCD.

In Table 2, we summarize our results for the $\eta$ photoproduction
showing the model-independent parameters we have extracted from
the data.
There is a substantial disagreement with a quark model calculation of
these parameters in the approach of 
Capstick and Roberts\cite{ref15}. This is an urgent worry for
the theorists. However, the ratio $A_{1/2}^{n}/A_{1/2}^{p}$ is
predicted correctly in the quark model: we get
\begin{equation}
A^{n}_{1/2}/A_{1/2}^{p}=-0.84\pm 0.15,
\end{equation}
and the quark models\cite{ref15} give $-0.83$. In this ratio,
the strong interaction physics drops. What does this apparent
agreement and the disagreement in Table 2 mean? We need to 
understand this better.
\begin{table}[t]
\protect
\tcaption{Proton and neutron electrostrong parameters $\xi_{p}$,
$\xi_{n}$ for N$^*$(1535) excitation and decay via the eta
channel. Results are all in units of $10^{-4}$ MeV$^{-1}$.}
\small
\vspace{0.4cm}
\begin{center}
\begin{tabular}{|c|c|c|}
\hline
& &  \\
&Our result\protect\cite{ref3} & Quark model\protect\cite{ref15}
\\ \hline
$\xi_{p}$& $2.20\pm 0.15$& 1.13\\ \hline
$\xi_{n}$& $-1.86\pm 0.20$ &$-0.94$\\
& &   \\ \hline
\end{tabular}
\end{center}
\end{table}

We shall return to the emerging results on polarization observables in
future communications. Suffice to say that the polarization 
obsevables will tell us more about the role of N$^*$(1520)\cite{ref28}.
 
\section{Our Results for Electroproduction}\label{sec:elec}
Here we must make use of the old data base\cite{ref9}, since
precious few new results are in. Our results are best sumarized in
Fig.3, where we plot the parameter $\xi_{T}$, extracted from the
existing data, as a function of $Q^2$. The important result here is the 
{\it utter failure} of the quark model to reproduce this nearly
 model-independent parameter, extracted from the data by us\cite{ref4}.
This figure, along with Table 2, constitute our main result.
\begin{figure}[t]
\begin{center}
\psfig{figure=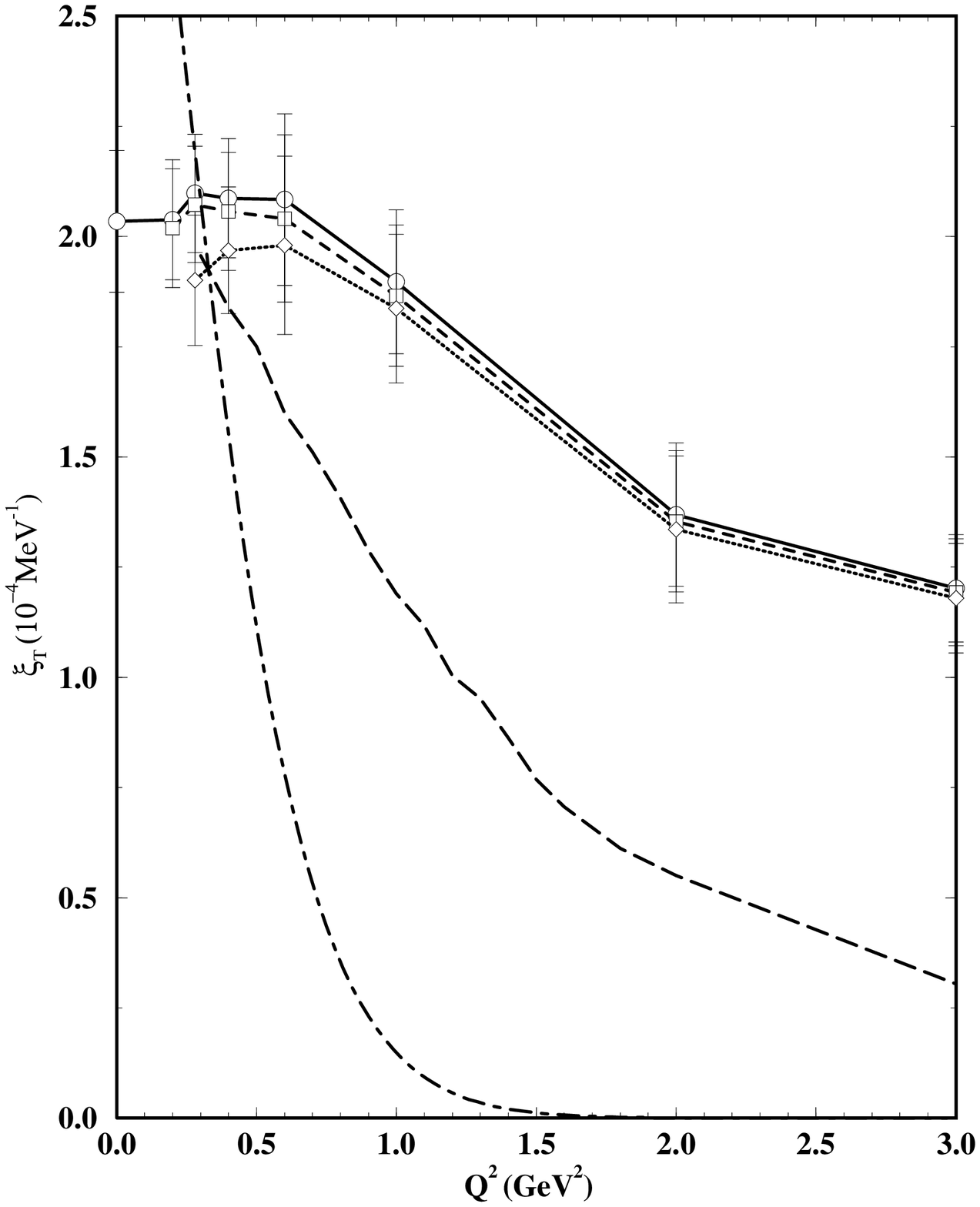,height=3.5in}
\end{center}
\fcaption{$\xi_{T}$ vs. $Q^2$ for different prescriptions\protect\cite{ref4}
of $S_{1/2}$ to $A_{1/2}$ ratio: (a) set $S_{1/2}=0$ (circles
connected by a solid line); (b) fix $S_{1/2}/A_{1/2}$ by the
quark shell model\protect\cite{ref15} (squares connected by a dashed line);
(c) use the value of $S_{1/2}$ from refs. [14, 15, 16] of
ref. [29] (diamonds connected by a dotted line). The non-relativistic
quark model prediction of ref.15 is the dot-dashed line. The prediction 
from a light front approach of Stanley and Weber\cite{ref29} is
also shown (long-dashed line), with their parameter $\alpha =0.2
GeV^2$.
\label{fig:fig3}}
\end{figure}

\section{Conclusions}
Paul Stoler has asked\cite{ref29} a question earlier during this workshop:
Would our current effort be written up in the {\it New York Times}?
We know the answer to that one! Neverthless, something very
exciting is happening right before us! For the N$^*$(1535), there is
{\it an acute shortage of the transition electricity in the
quark model}, compared to our observations at the photon point.
This shortage becomes even more serious at higher $Q^2$. This is
akin to the shortage of transition magnetism in the $N\rightarrow \Delta$
transition, where the quark model is also in serious trouble.
Thus, {\it history is being made, as the venerable quark model is failing
badlly} right before our eyes!

{\bf QCD, come and rescue us!}
 
\section{Acknowledgments}
One of us (NCM) thanks the organizers for the invitation. NCM and MB
thank the INT for the wonderful hospitality. NCM
has the special pleasure to thank Harry Lee for
many stimulating conversations. We thank R. Davidson
for many discussions. The research of NCM and
JFZ is supported in part by the U.S. Department of Energy. The research
of MB is supported by the Natural Sciences and Engineering Research Council 
of Canada.

\end{document}